\documentclass[aps,prl, twocolumn, superscriptaddress, longbibliography]{revtex4-1}

\usepackage{graphicx}
\usepackage{braket}

\usepackage{hyperref}
\usepackage{amsmath}

\usepackage{float}

\begin{document}


\title{Inverted orbital polarization in strained correlated oxide films}


\author{Paul C. Rogge}
\email[]{progge@drexel.edu}
\affiliation{Department of Materials Science and Engineering, Drexel University, Philadelphia, Pennsylvania 19104, USA}

\author{Robert J. Green}
\affiliation{Stewart Blusson Quantum Matter Institute, University of British Columbia, Vancouver, British Columbia V6T 1Z4, Canada}
\affiliation{Department of Physics \& Engineering Physics, University of Saskatchewan, Saskatoon, Saskatchewan S7N 5E2, Canada}

\author{Padraic Shafer}
\affiliation{Advanced Light Source, Lawrence Berkeley National Laboratory, Berkeley, California 94720, USA}

\author{Gilberto Fabbris}
\affiliation{Department of Condensed Matter Physics and Materials Science, Brookhaven National Laboratory, Upton, New York 11973, USA}

\author{Andi M. Barbour}
\affiliation{National Synchrotron Light Source II, Brookhaven National Laboratory, Upton, New York 11967, USA}

\author{Benjamin M. Lefler}
\affiliation{Department of Materials Science and Engineering, Drexel University, Philadelphia, Pennsylvania 19104, USA}

\author{Elke Arenholz}
\affiliation{Advanced Light Source, Lawrence Berkeley National Laboratory, Berkeley, California 94720, USA}

\author{Mark P. M. Dean}
\affiliation{Department of Condensed Matter Physics and Materials Science, Brookhaven National Laboratory, Upton, New York 11973, USA}

\author{Steven J. May}
\email[]{smay@coe.drexel.edu}
\affiliation{Department of Materials Science and Engineering, Drexel University, Philadelphia, Pennsylvania 19104, USA}


\date{\today}

\begin{abstract}
Manipulating the orbital occupation of valence electrons via epitaxial strain in an effort to induce new functional properties requires considerations of how changes in the local bonding environment affect the band structure at the Fermi level. Using synchrotron radiation to measure the x-ray linear dichroism of epitaxially strained films of the correlated oxide CaFeO\textsubscript{3}, we demonstrate that the orbital polarization of the Fe valence electrons is opposite from conventional understanding. Although the energetic ordering of the Fe $3d$ orbitals is confirmed by multiplet ligand field theory analysis to be consistent with previously reported strain-induced behavior, we find that the nominally higher energy orbital is more populated than the lower. We ascribe this inverted orbital polarization to an anisotropic bandwidth response to strain in a compound with nearly filled bands. These findings provide an important counterexample to the traditional understanding of strain-induced orbital polarization and reveal a new method to engineer otherwise unachievable orbital occupations in correlated oxides. 
\end{abstract}


\maketitle


The use of epitaxial strain to induce occupation of specific electron orbitals by removing orbital degeneracies has been pursued in transition metal oxides in an effort to engineer new electronic and magnetic properties \cite{Tokura_Manganites_XLD, Aruta_LSMO_XLD, Strain_OP_Csiszar, Hansmann_Held_theory_nickelate_SL_OP, Nickelate_holes_reduced_OP, Chak_asymmetric_XLD, Freeland_LNO_XLD, Wu_nickelate_SL_XLD, Wu_PrNiO3_orbital_polarization, Bruno_nickelate_XLD, Pesquera_LSMO_XLD}. Such strain-induced orbital polarization has been very successfully described by ligand field theory, which considers the overlap of electron orbitals between a central cation and its surrounding anions \cite{Cox_oxide_book, Khomskii_book}. For transition metal perovskite oxides, the metal cation is octahedrally coordinated by six oxygen anions, or ligands. This $O_h$ symmetry splits the five degenerate $d$-levels into two groups: a lower, triply degenerate group ($t_{2g}$) and a doubly degenerate group ($e_g$) higher in energy by an amount $10Dq$. Whereas the lobes of the O $p$ orbitals point in between the $t_{2g}$ lobes, they directly overlap with the $e_g$ lobes, which comes at a coulombic energy cost that raises the $e_g$ orbitals in energy. Epitaxial strain alters the local crystal field and lifts the $t_{2g}$ and $e_g$ degeneracies. For example, tensile strain reduces the overlap between the $e_g$ orbital of $d_{x^2-y^2}$ symmetry and its ligands, thus lowering its energy relative to the other $e_g$ orbital, $d_{3z^2-r^2}$, by an amount $\Delta e_g$ [see Fig. 1(a) inset]. Unless the $e_g$ orbitals are fully filled, one subsequently expects $d_{x^2-y^2}$ to be preferentially occupied; the converse applies for compressive strain. This simple picture has been used to explain strain-induced orbital polarization in many systems, particularly \textit{AB}O\textsubscript{3} perovskite oxides \cite{Tokura_Manganites_XLD, Aruta_LSMO_XLD, Hansmann_Held_theory_nickelate_SL_OP, Nickelate_holes_reduced_OP, Chak_asymmetric_XLD, Freeland_LNO_XLD, Wu_nickelate_SL_XLD, Wu_PrNiO3_orbital_polarization, Bruno_nickelate_XLD, Pesquera_LSMO_XLD}. In this Letter, we find that this model fails to explain orbital polarization in strained films of CaFeO\textsubscript{3}, which exhibit orbital polarization opposite to that described above. 

To quantify the electron occupation of specific $e_g$ orbitals, we measure x-ray absorption across the Fe $L$- and O $K$-edge resonance energies using linearly polarized photons, which allows us to differentiate between $d_{x^2-y^2}$ and $d_{3z^2-r^2}$ occupations. Analyzing the x-ray linear dichroism using multiplet ligand field theory reveals that the effect of epitaxial strain on the energetic ordering of the $e_g$ orbitals is consistent with the aforementioned considerations--stretched bonds are lower in energy than unstretched. Given this energetic landscape, however, the expected orbital \textit{occupations} do not follow: The out-of-plane ($d_{3z^2-r^2}$) orbitals are more populated under tensile strain (and \textit{vice versa} for compressive strain). We propose that this inverted orbital polarization arises from strain-induced anisotropic changes in the Fe-O-Fe bond angles and the resulting anisotropic bandwidths in bands that are more-than-half-filled. Such conditions are not limited to ferrates but could arise in other strongly hybridized systems, such as the rare-earth nickelates \cite{Sawatzky_bond_disproportionation}.

 \begin{figure}
 \includegraphics{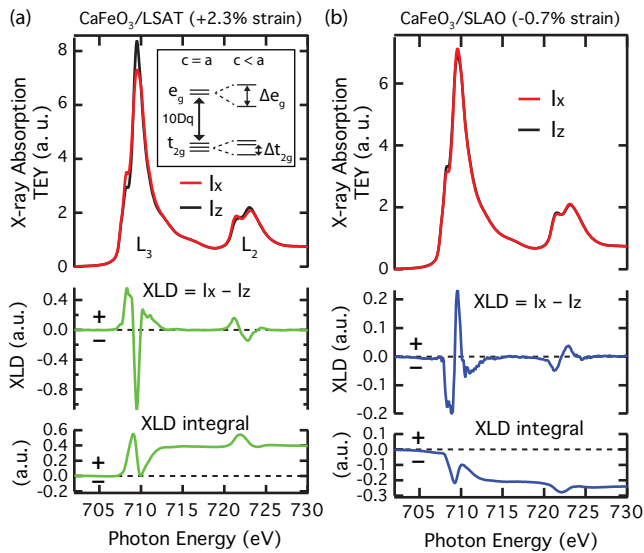}
 \caption{Polarization-dependent x-ray absorption measured by total electron yield (TEY) across the Fe $L$-edge for CaFeO\textsubscript{3} under (a) tensile strain and (b) compressive strain. Inset:  Octahedral crystal field splitting of transition metal $d$ levels for a (001)-oriented film under no strain ($c=a$) and under biaxial tensile strain ($c<a$). \label{Fig1_Fe_XLD}}
 \end{figure}

CaFeO\textsubscript{3} films of 40 pseudocubic unit cells (${\sim}$15 nm thick) were deposited by oxygen plasma-assisted molecular beam epitaxy. Epitaxial strain was achieved by deposition on single crystal, (001)-oriented substrates: YAlO$_3$ (YAO, -2.0\% strain), SrLaAlO$_4$ (SLAO, -0.7\%), LaAlO$_3$ (LAO, 0.2\%), (La$_{0.18}$Sr$_{0.82}$)(Al$_{0.59}$Ta$_{0.41}$)O$_3$ (LSAT, 2.3\%), and SrTiO$_3$ (STO, 3.3\%). As previously reported, the films are coherently strained and exhibit bulk-like electrical transport, indicating high-quality, stoichiometric films \cite{Rogge_PRM}. Prior to all measurements, the films were reoxidized by heating to ${\sim}$600 $^\circ$C in oxygen plasma (200 Watts, 1x$10^{-5}$ Torr chamber pressure) and then slowly cooled to room temperature in oxygen plasma. X-ray absorption spectroscopy was performed at the Advanced Light Source, Beamline 4.0.2 and at the National Synchrotron Light Source-II, Beamline 23-ID-1. The spectra were recorded at 290 K, where CaFeO\textsubscript{3} is paramagnetic with metallic conductivity \cite{Woodward_CFO}. The x-ray incident angle was $20^{\circ}$ from the film plane, and a geometric correction was applied to the absorption measured with photons polarized out of the film plane \cite{SI_XLD}. 

Although CaFeO\textsubscript{3} has an unusually high formal oxidation state of Fe\textsuperscript{4+}, its ground state exhibits a significant self-doped ligand hole density due to its negative charge transfer energy, $\Delta$ \cite{Kawasaki_CFO_first_transport, Bocquet_SFO_ligand_holes, Woodward_CFO, Matsuno_CFO_dispro, Takeda_CFO, Rogge_PRM}. In this regime, the transition metal cation does not adopt its formal oxidation state but instead keeps an extra electron that results in a hole ($\underline{L}^1$) on the oxygen ligand \cite{ZSA, Sawatzky_neg_charge_trans_1, Matsuno_CFO_dispro}. So while CaFeO\textsubscript{3} has a nominal Fe configuration of $d^4$ ($e_g^1$), its ground state has a strong $d^5\underline{L}^1$ contribution. Because of the half-filled $d$-shell, this $d^5\underline{L}^1$ ($e_g^2$) state has no significant orbital polarization and is expected to decrease the degree of orbital polarization achievable in the Fe states.

X-ray absorption across the Fe $L$-edge for a CaFeO\textsubscript{3} film under tensile strain is shown in Fig. \ref{Fig1_Fe_XLD}(a). The $L_3$ peak exhibits primarily a single, broad peak (with a small shoulder) that is consistent with nominal Fe\textsuperscript{4+} \cite{Abbate_SFO_XAS, Reduced_SFO_XAS} and significantly contrasts with the well-separated double peak structure seen in Fe\textsuperscript{3+} perovskites, such as LaFeO\textsubscript{3} and EuFeO\textsubscript{3} \cite{Reduced_SFO_XAS, Amber_EuFeO3}. This spectral signature as well as the bulk-like electrical transport indicate that oxygen vacancies have been sufficiently suppressed. As seen in Fig. \ref{Fig1_Fe_XLD}(a), the x-ray absorption is polarization-dependent. The difference in absorption measured with photons polarized parallel to the film plane, $I_x$, and photons polarized out of the film plane, $I_z$, is termed x-ray linear dichroism (XLD = $I_x-I_z$). The XLD shows areas of both positive and negative intensity, and this lineshape is similar but nearly opposite in sign for the compressively strained film, CaFeO\textsubscript{3}/SLAO, shown in Fig. \ref{Fig1_Fe_XLD}(b). Because $I_x$ preferentially probes empty states in $d_{x^2-y^2}$ and $I_z$ probes $d_{3z^2-r^2}$, their difference in total integrated intensity is a measure of the orbital polarization \cite{Thole_dichroism_prediction, van_der_Laan_XLD}, and indeed the XLD integrals are non-zero.

Evaluating the sign of the integrated XLD, however, uncovers a surprising result: the $e_g$ electron occupation does not follow the conventional ligand field model. For tensile strain the positive XLD integral implies more empty $d_{x^2-y^2}$ states. Thus under tensile strain CaFeO\textsubscript{3} has more electrons in $d_{3z^2-r^2}$, which is opposite of that predicted by ligand field theory. Under compressive strain, the integrated XLD sign implies that $d_{x^2-y^2}$ has more electrons. This behavior is consistent among the other films: The integrated XLD for tensile CaFeO\textsubscript{3} on STO (+3.3\% strain) is positive, compressed CaFeO\textsubscript{3} on YAO (-2.0\%) is negative, and the relatively unstrained CaFeO\textsubscript{3} film on LAO (+0.2\%) is approximately zero \cite{SI_XLD}. 

\begin{figure}
 \includegraphics{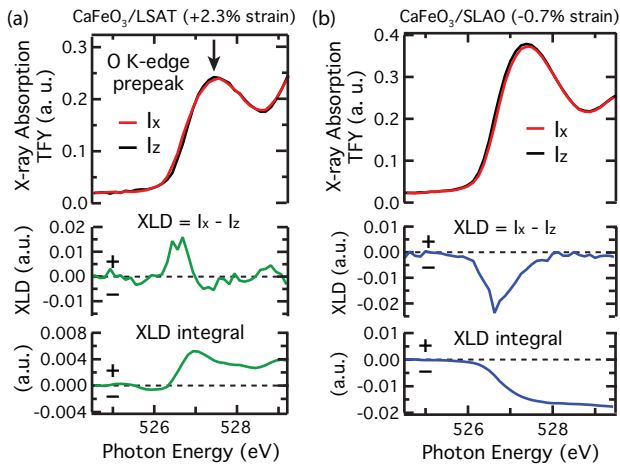}
 \caption{Polarization-dependent x-ray absorption of the O $K$-edge prepeak (arrow) for CaFeO\textsubscript{3} under (a) tensile and (b) compressive strain measured by total fluorescence yield. \label{O_XAS}}
 \end{figure}

In order to verify these relative $e_g$ occupations, we repeated the XLD measurements at the O $K$-edge. This transition probes unoccupied states with O $2p$ character, which are strongly hybridized with Fe $3d$ states due to the negative charge transfer energy \cite{Abbate_SFO_XAS}. Because these ligand states have the same symmetry as the Fe $3d$ states that they hybridize with \cite{Sawatzky_bond_disproportionation}, they are expected to mimic the Fe $e_g$ occupation. We particularly focus on the O $K$-edge prepeak feature between 526 and 529 eV because it directly probes the oxygen ligand hole states \cite{Abbate_SFO_XAS, Chen_cuprate_O_prepeak, Suntivich_O_Kedge_holes, Pellegrin_holes_prepeak}. We note that oxygen in the substrates contributes only at energies above the prepeak. As seen in Fig. \ref{O_XAS}, the oxygen prepeak exhibits linear dichroism, where the tensile strained film, CaFeO\textsubscript{3}/LSAT, has positive dichroism and the compressively strained film, CaFeO\textsubscript{3}/SLAO, has negative dichroism. A positive integrated XLD indicates more empty states in the $p_x$ and $p_y$ orbitals compared to $p_z$--that is, under tensile strain, more electrons have $p_z$ character than $p_x$ and $p_y$; the opposite situation exists for the film under compressive strain. This precisely mirrors the $e_g$ occupation measured for the Fe $3d$ states. The other strained films (CaFeO\textsubscript{3}/STO, CaFeO\textsubscript{3}/LAO, CaFeO\textsubscript{3}/YAO) exhibit an O prepeak XLD consistent with the two films highlighted here \cite{SI_XLD}. 
 
With the qualitative occupation of the Fe $e_g$ orbitals confirmed, we now quantitatively estimate the orbital polarization by computing the hole ratio, $\int I_x dE$/$\int I_z dE$ \cite{Thole_hole_ratio, Haverkort_dissertation}. The hole ratio depends on the Fe valence filling, and for high-spin CaFeO\textsubscript{3} there are three $t_{2g}$ electrons and, as will be shown below, we find that the total $e_g$ occupation is 1.85 electrons. For CaFeO\textsubscript{3}/LSAT (tensile), the hole ratio is 1.018, which gives 0.90 electrons in $d_{x^2-y^2}$ and 0.95 electrons in $d_{3z^2-r^2}$ \cite{SI_XLD}, or an orbital polarization of ${\sim}$6\%. Repeating for compressively strained CaFeO\textsubscript{3}/SLAO, we find 0.93 electrons in $d_{x^2-y^2}$ and 0.91 electrons in $d_{3z^2-r^2}$, or ${\sim}$2\% polarized. This smaller orbital polarization is consistent with its lower strain state (-0.7\%) compared to CaFeO\textsubscript{3}/LSAT (+2.3\%). 

To help interpret these findings, we analyze the x-ray absorption for CaFeO\textsubscript{3} using multiplet ligand field theory of a FeO\textsubscript{6} cluster \cite{Groot_XAS_multiplet_review}. We begin with the formal Fe$^{4+}$ ($3d^4$) occupation and full ligand orbitals while including a negative charge transfer energy \cite{Rogge_PRM} such that the configuration interaction ground state has primarily a $d^5\underline{L}^1$ character but still exhibits the same $S=2$ high spin symmetry of the $3d^4$ case \cite{Sawatzky_bond_disproportionation}. This $S=2$ configuration has a two-fold degeneracy due to the hybridized $e_g$ orbitals that is lifted by the imposed strain (corresponding to preferential occupation of $d_{x^2-y^2}$ and preferential occupation of $d_{3z^2-r^2}$, respectively). Further, each of these $S=2$ states has a five-fold spin degeneracy, which is lifted via the atomic spin-orbit interaction and non-tetragonal local crystal field distortions. We neglect the latter and hence label the spin-orbit split states by $J_z = 0,\pm 1,\pm 2$. Thus our XAS and XLD spectra are expected to be linear combinations of two sets of 5 spectra, one set corresponding to preferential $d_{x^2-y^2}$ occupation and one for preferential $d_{3z^2-r^2}$ occupation \cite{Wu_nickelate_SL_XLD}. The model parameters were optimized by comparing the calculated XLD to the experimental XLD \cite{SI_XLD}. 

\begin{figure*}
 \includegraphics{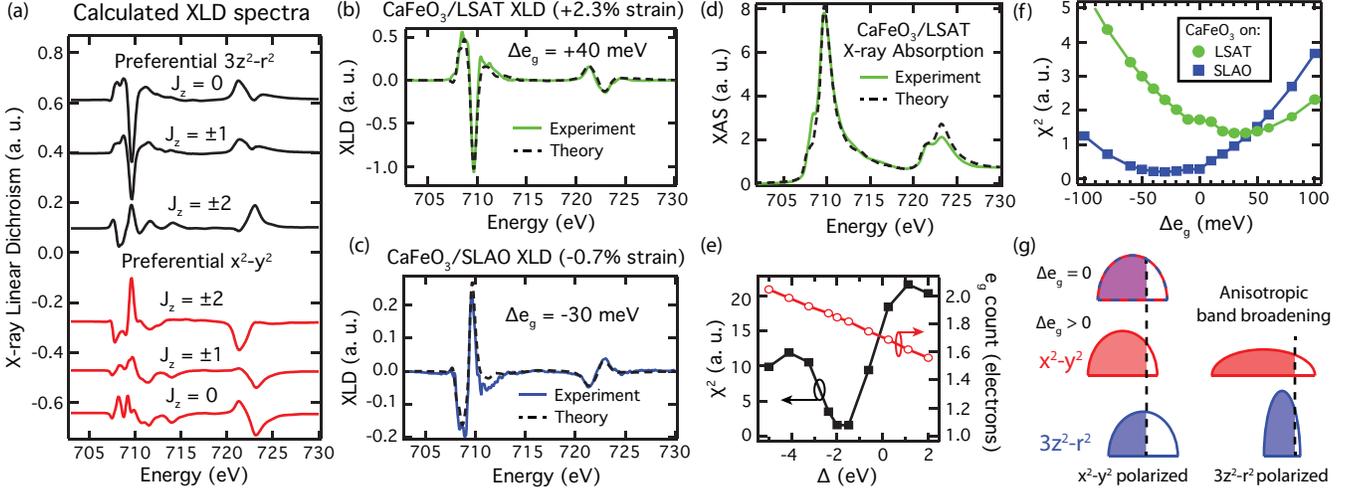}
 \caption{(a) Linear dichroism spectra calculated using multiplet ligand field theory for a FeO$_6$ cluster with preferential $d_{x^2-y^2}$ occupation and preferential $d_{3z^2-r^2}$ occupation under moderate tensile strain ($\Delta e_g = +40$ meV; $\Delta = -2.0$ eV). Because the $J_z$ doublets exhibit nearly identical spectra [\textit{i.e.}, $(J_z = +1) \equiv (J_z = -1$)], their averaged spectrum is shown, reducing the number of XLD spectra from 10 to six. A combination of the calculated XLD spectra were fit to the experimental XLD for CaFeO\textsubscript{3} under (b) tensile and (c) compressive strain. (d) From the best XLD fit for CaFeO\textsubscript{3}/LSAT, the resulting x-ray absorption spectrum is compared to experiment. (e) The $\chi^2$ value (filled squares) for the $J_z$ fit to CaFeO\textsubscript{3}/LSAT is minimized for negative values of $\Delta$. The total number of $e_g$ electrons (open circles) increases as $\Delta$ decreases. (f) The best XLD fit occurs for $\Delta e_g > 0$ for tensile strained CaFeO\textsubscript{3}/LSAT and for $\Delta e_g < 0$ for compressively strained CaFeO\textsubscript{3}/SLAO. (g) Simplified schematic of the proposed effect of anisotropic bandwidths on the resulting orbital polarization under tensile strain ($\Delta e_g>0$) for a band with greater than half-filling (Fermi level indicated by the vertical, dashed line; filled states are shaded). \label{Fig_Jz}}
 \end{figure*}

The XLD from these two sets of five calculated spectra are shown in Fig. \ref{Fig_Jz}(a) for moderate tensile strain ($\Delta e_g = +40$ meV). At finite temperature, the experimental spectrum is expected to be a combination of these $J_z$ spectra \cite{Haverkort_LaTiO3_Jz_boltzmann}, depending on their relative energies due to the spin-orbit splitting and low symmetry crystal field distortions. Therefore, a least-squares fitting procedure was used to determine a coefficient value for each of the $J_z$ XLD spectra such that the resulting combination produces the best fit with experiment. This calculated XLD spectrum has a corresponding x-ray absorption spectrum, and the XLD fitting was constrained such that the resulting calculated x-ray absorption spectral weight ($I_x + I_z$) is within $\pm1$\% of the experimental spectral weight. 

As seen in Figs. \ref{Fig_Jz}(b) and \ref{Fig_Jz}(c), the experimental XLD is well-captured by the $J_z$ fit for both CaFeO\textsubscript{3}/LSAT and CaFeO\textsubscript{3}/SLAO. All major features of the $L_3$ and $L_2$ XLD peaks are replicated. The corresponding x-ray absorption spectrum of the optimized XLD fit for CaFeO\textsubscript{3}/LSAT, shown in Fig. \ref{Fig_Jz}(d), also has excellent agreement with experiment. The coefficients for each $J_z$ spectrum are listed in the Supplemental Material \cite{SI_XLD}. Fig. \ref{Fig_Jz}(e) highlights that the goodness of fit, $\chi^2$, is a strong function of $\Delta$, and the lowest $\chi^2$ values are obtained for $\Delta < 0$, further confirming that CaFeO\textsubscript{3} is a negative charge transfer material. We find that $\Delta = -2.0$ eV provides the best fit to experiment, which is in good agreement with previously reported values for formal Fe\textsuperscript{4+} SrFeO\textsubscript{3} \cite{Bocquet_SFO_ligand_holes}, and is more negative than the rare-earth nickelates \cite{Wu_nickelate_SL_XLD, Robert} but not so negative that the $t_{2g}$ and $e_g$ levels are inverted, as in some compounds \cite{Ushakov_crystal_field}. This value sets the number of self-doped ligand holes, and as seen in Fig. \ref{Fig_Jz}(e), for $\Delta = -2.0$ eV the Fe $e_g$ occupation is 1.85 electrons. This large Fe $e_g$ occupation is consistent with the small measured Fe $3d$ orbital polarization. 

The XLD fits also reproduce the measured Fe orbital polarization. Converting the preferential $x^2-y^2$ and the preferential $3z^2-r^2$ fit contributions to a $d_{x^2-y^2}$ and $d_{3z^2-r^2}$ occupation, we find that the orbital occupation for tensile CaFeO\textsubscript{3}/LSAT exhibits a small preference for $d_{3z^2-r^2}$, where $d_{x^2-y^2}$ has 0.91 electrons and $d_{3z^2-r^2}$ has 0.94 electrons \cite{SI_XLD}. This difference of 0.03 electrons agrees well with the difference of 0.05 electrons determined by the sum rule analysis of the XLD integrals. For the compressively strained film, CaFeO\textsubscript{3}/SLAO, the best $J_z$ fit is with equal occupation of 0.93 electrons in both $d_{x^2-y^2}$ and $d_{3z^2-r^2}$. 

Of particular note is the sign of the strain-induced crystal field $e_g$ splitting, $\Delta e_g$, that produces the best agreement with experiment. As seen in Fig. \ref{Fig_Jz}(f), for tensile strained CaFeO\textsubscript{3}/LSAT, the lowest $\chi^2$ value occurs for +40 meV; for compressively strained CaFeO\textsubscript{3}/SLAO, -30 meV produces the best fit. These magnitudes are of the same order as other similarly strained perovskite oxides \cite{Aruta_LSMO_XLD, Wu_nickelate_SL_XLD, Gilberto_Nickelate_XLD}. Importantly, the respective signs indicate an $e_g$ splitting consistent with the traditional ligand field model: $\Delta e_g > 0$ implies that $d_{x^2-y^2}$ is lower in energy than $d_{3z^2-r^2}$, which would be expected for tensile strain, and \textit{vice versa} for compressive strain. This provides a critical insight: The energetic landscape of the Fe $3d$ orbitals follows the typical ligand field understanding, where, for example, tensile strain lowers $d_{x^2-y^2}$ in energy relative to $d_{3z^2-r^2}$. Despite this, the $d_{x^2-y^2}$ orbital has fewer electrons than $d_{3z^2-r^2}$ in the film under tensile strain, indicating an inversion in orbital polarization.

What, then, overrides the $\Delta e_g$ splitting and produces the inverted $e_g$ orbital occupation? Although oxygen vacancies can be equatorially or apically ordered under epitaxial strain \cite{Spaldin_O_vacancy_ordering}, the resulting preferential orbital occupation would be opposite of the results here. Moreover, because our experimental findings are not replicated by previous density functional theory calculations \cite{Antonio_CFO_strain}, we propose a new mechanism. It is well known that perovskites can accommodate epitaxial strain by changes in both bond lengths and rotations of the octahedral complexes surrounding the transition metal (TM) cation \cite{Miniotas_strain_rotations, Xie_strain_rotations, Steve_octahedral_rotations_strain, Rondinelli_Spaldin_Adv_Mater}. Rotations alter the TM-O-TM bond angle, and angles less than $180^{\circ}$ have reduced orbital overlap and thus narrower bands. For a perovskite that exhibits rotations in its bulk form, such as CaFeO\textsubscript{3}, biaxial tensile strain increases the in-plane TM-O-TM bond angle towards $180^{\circ}$, whereas the out-of-plane angle decreases further and is typically more strongly affected than the in-plane angles \cite{Rondinelli_Spaldin_Adv_Mater, Antonio_CFO_strain}. Thus in the simplest approximation where strain is accommodated predominantly by octahedral rotations, under tensile strain one would expect the in-plane (x, y) bandwidth to increase and the out-of-plane (z) bandwidth to decrease. 

Such anisotropic bandwidth effects can lead to an inverted orbital polarization in compounds with greater-than-half-filled bands. As illustrated in Fig. \ref{Fig_Jz}(g) for the case under tensile strain, $\Delta e_g > 0$ shifts the band center of masses, but a broadening of the $x^2-y^2$ band and a narrowing of the $3z^2-r^2$ band can result in $3z^2-r^2$ being more occupied than $x^2-y^2$. The precise orbital polarization is expected to depend on the specific bandstructure and Fermi level position. For bands with half-filling or less, the same anisotropic bandwidths result in the conventional orbital polarization and thus do not replicate our findings \cite{SI_XLD}. We further note that this effect does not require metallicity and indeed when repeating the Fe $L$-edge XLD measurements at lower temperatures (180 K) where CaFeO\textsubscript{3} is insulating, the inverted orbital polarization is maintained \cite{SI_XLD}. 

In summary, we have shown that epitaxially strained films of CaFeO\textsubscript{3} exhibit orbital polarization that responds to the strain state in a way that requires considerations beyond the commonly assumed ligand field model. By analyzing the x-ray linear dichroism with multiplet ligand field simulations, we find that under tensile strain the $e_g$ electronic population is weighted towards $d_{3z^2-r^2}$ orbitals, despite being ${\sim}$40 meV higher in energy than $d_{x^2-y^2}$. The opposite is observed under compressive strain. We propose an explanation for this behavior by considering anisotropic modifications of the bandwidth of the $e_g$ states, in which under tensile strain a broadened $d_{x^2-y^2}$ band and a narrowed $d_{3z^2-r^2}$ lead to this inverted orbital polarization configuration. This scenario is consistent with the orbital energetic landscape as determined by ligand field theory, as well as the measured film strain, under the assumption that strain is accommodated primarily by octahedral bond rotations. More generally, our results demonstrate that effects typically not considered in the conventional understanding of strain-induced orbital polarization can mitigate or even invert the orbital polarization. This highlights that the interpretation of orbital polarization in ultrathin films and short-period superlattices \cite{Chak_asymmetric_XLD, Freeland_LNO_XLD, Keimer_Nickelate_reflectometry, Cao_XLD, Wu_PrNiO3_orbital_polarization, Wu_nickelate_SL_XLD, Disa_Ahn_tricolor_SL}, where non-bulk octahedral rotations can be induced, should include such considerations. Additionally, these results demonstrate that bandwidth control is a potentially new way to engineer orbital polarization in correlated oxides. 

\begin{acknowledgments}
We thank G. Sawatzky and A. Fujimori for helpful discussions. PCR and SJM were supported by the Army Research Office, grant number W911NF-15-1-0133, and film synthesis at Drexel utilized deposition instrumentation acquired through an Army Research Office DURIP grant (W911NF-14-1-0493). RJG was supported by the Natural Sciences and Engineering Research Council of Canada. Work at Brookhaven National Laboratory was supported by the U.S. Department of Energy (DOE), Office of Basic Energy Sciences under Contract No. DE-SC0012704 and Early Career Award Program under Award No. 1047478. This work used resources at the Advanced Light Source, which is a DOE Office of Science User Facility under contract No. DE-AC02-05CH11231, and at Beamline 23-ID-1 of the National Synchrotron Light Source II, a DOE Office of Science User Facility operated for the DOE Office of Science by Brookhaven National Laboratory under Contract No. DE-SC0012704.
\end{acknowledgments}


%


\clearpage
\onecolumngrid
\begin{center}
\textbf{\large Supplemental Material: Inverted orbital polarization in strained correlated oxide films}
\end{center}

\subsection{I. X-ray absorption experiment and additional data}

 \begin{figure}[H]
 \includegraphics{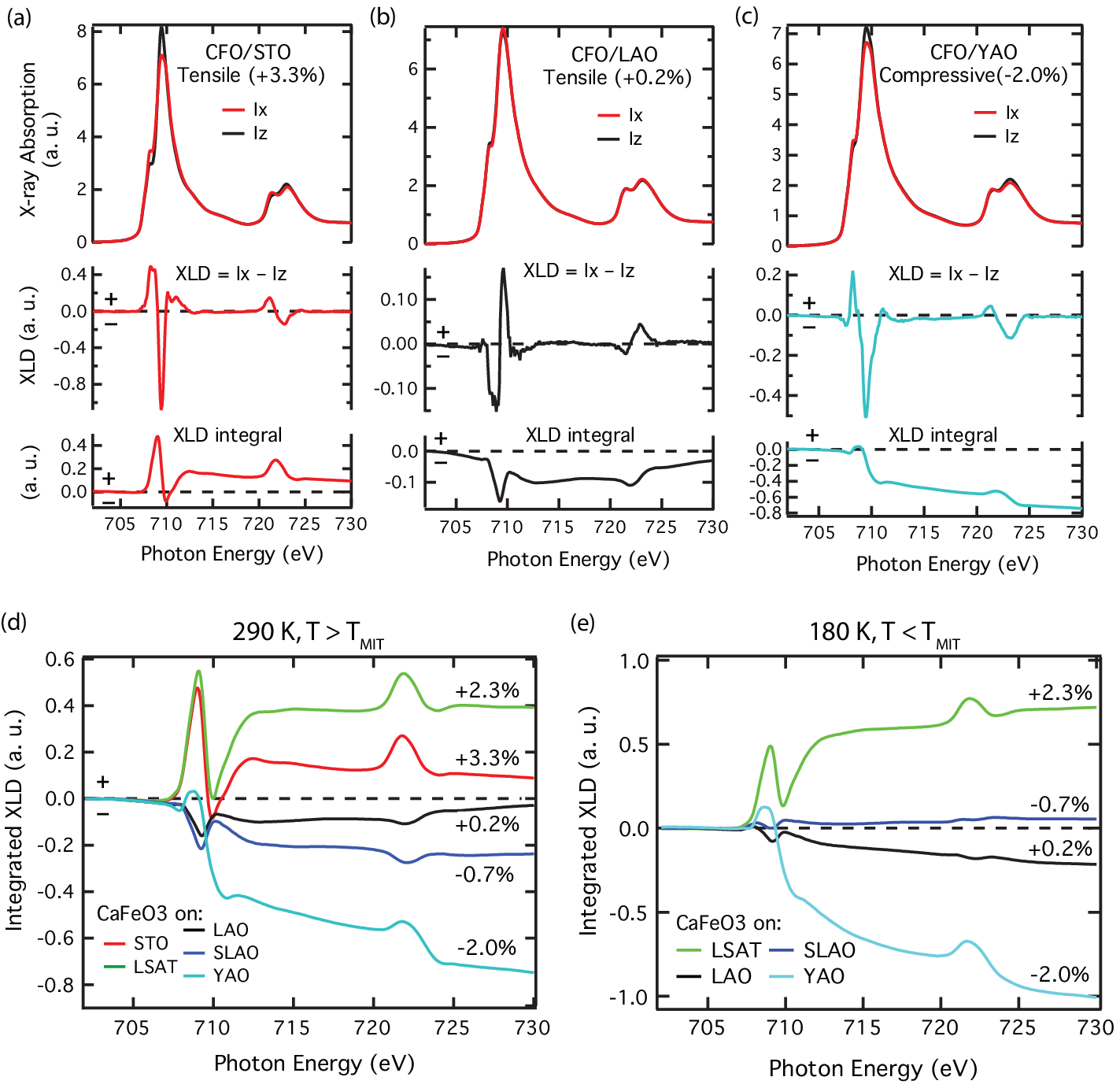}
 \centering
 \caption{Fe $L$-edge x-ray absorption data of additional CaFeO\textsubscript{3} films. Polarization-dependent x-ray absorption across the Fe $L$-edge for (a) tensile (+3.3\%) CaFeO\textsubscript{3} on SrTiO\textsubscript{3}, (b) relatively unstrained CaFeO\textsubscript{3} on LaAlO\textsubscript{3} (+0.2\%), and (c) compressively strained (-2.0\%) CaFeO\textsubscript{3} on YAlO\textsubscript{3}. The resulting x-ray linear dichroism ($I_x - I_z$) and its integral are also shown. Measurements were made above CaFeO\textsubscript{3}'s metal-insulator transition temperature (\textit{T}\textsubscript{MIT}). (d) The integrated XLD of all CaFeO\textsubscript{3} films for $T>$\textit{T}\textsubscript{MIT} (metallic) show consistent behavior: Tensile strained films have a positive XLD integral and compressively strained films have a negative XLD integral. (e) This behavior is maintained in the insulating state ($T<$\textit{T}\textsubscript{MIT}). Data for CaFeO\textsubscript{3}/STO sample was not obtained at 180 K. 
 \label{SI_Fe_XAS}}
 \end{figure}
 
\begin{figure}[H]
 \centering
 \includegraphics{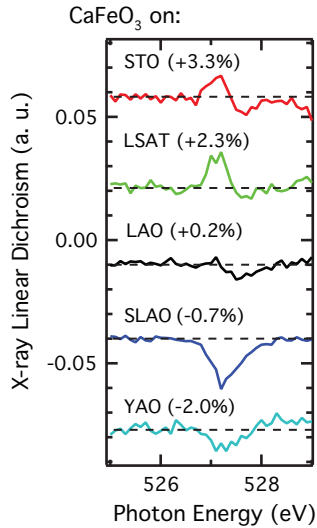}
 \caption{O $K$-edge prepeak x-ray linear dichroism for all CaFeO\textsubscript{3} films measured by total fluorescence yield. The strain-dependent behavior is consistent with that seen for the Fe $3d$ $e_g$ occupation.
 \label{SI_O_XAS}}
 \end{figure}

 The x-ray incident angle was $20^{\circ}$ relative to the film plane and the polarization was controlled upstream. A geometric correction was applied to the absorption measured with photons polarized out of the film plane, $I_{\pi}$:
\begin{equation}
I_z = \frac{(I_\pi - I_x\sin^2(\theta))}{\cos^2(\theta)},
\end{equation}
where $I_x$ is the absorption intensity measured with photons polarized parallel to the film plane. At least 12 scans of each polarization were performed for the Fe $L$-edge measurements ($680-750$ eV), and at least four scans for the O $K$-edge ($513-555$ eV). The spectra were normalized by setting the pre-edge intensity to zero by subtracting a line fit to the pre-edge, followed by setting the post-$L_2$ intensity to unity at 750 eV for the Fe scans and setting the maximum intensity to unity for the O scans. The Fe x-ray absorption and XLD of the additional samples are shown in Fig. \ref{SI_Fe_XAS}, and the O $K$-edge prepeak XLD of the additional samples are shown in Fig. \ref{SI_O_XAS}.

 \subsection{II. Hole ratio derivation for CaFeO\textsubscript{3}}
 Following Refs. \cite{Thole_hole_ratio, Haverkort_dissertation}, the relative intensities for $x$, $y$, and $z$ polarized light are given by
 \begin{equation}
\begin{split}
I_x &= \frac{1}{n}(\frac{1}{2}n_{xy} + \frac{1}{2}n_{xz} + \frac{1}{6}n_{z^2} + \frac{1}{2}n_{x^2-y^2}) \\
I_y &= \frac{1}{n}(\frac{1}{2}n_{xy} + \frac{1}{2}n_{yz} + \frac{1}{6}n_{z^2} + \frac{1}{2}n_{x^2-y^2}) \\
I_z &= \frac{1}{n}(\frac{1}{2}n_{xz} + \frac{1}{2}n_{yz} + \frac{2}{3}n_{z^2} )
\end{split}
\end{equation}
where $I_j$ is the normalized intensity along direction $j$, $n_i$ is the number of holes in orbital $i$, and $n$ is the total number of holes. For high-spin CaFeO\textsubscript{3}, the $t_{2g}$ shell is half-full, and under moderate strains we assume negligible polarization of the $t_{2g}$ orbital occupation, \textit{i.e.},
\begin{equation}
\begin{split}
n_{xy} &= 1\\
n_{xz} &= 1\\
n_{yz} &= 1,\\
\end{split}
\end{equation}
where we leave the $e_g$ occupation as unknown. This then gives 
 \begin{equation}
\frac{I_x}{I_z} = \frac{\frac{1}{2} + \frac{1}{2} + \frac{1}{6}n_{z^2} + \frac{1}{2}n_{x^2-y^2}}{\frac{1}{2} + \frac{1}{2} + \frac{2}{3}n_{z^2}} = \frac{6 + n_{z^2}+3n_{x^2-y^2}}{6 + 4n_{z^2}},
\label{eq_hole_ratio}
\end{equation}
where $n_{x^2-y^2}$ and $n_{z^2}$ are the number of holes in the $d_{x^2-y^2}$ and $d_{3z^2-r^2}$ orbitals, respectively. This equation can only be solved if the total $e_g$ occupation is known. As discussed in the main manuscript, we estimate that the total $e_g$ occupation for CaFeO\textsubscript{3} is 1.85. 

\newpage
\subsection{III. Optimization of the ligand field model parameters}
We employed a standard multiplet ligand field theory model using the code \emph{Quanty} \cite{Haverkort_Wannier_PRB_2010, QuantyWeb}, which computes the eigenstates and spectra using exact diagonalization. The model includes the Fe $3d$ shell, a ligand shell comprised of $d$-symmetry linear combinations of oxygen $2p$ orbitals, and the Fe core $2p$ shell (needed for the spectroscopy simulations). Parameters of the model include the local Fe Coulomb and exchange integrals ($F^k_{dd}$, $F^k_{pd}$, and $G^k_{pd}$), for which we use Hartree-Fock determined values \cite{Cowan_TASS_1981} that are subsequently rescaled to account for atomic as well as solid state corrections using inter- and intra-shell rescaling factors $\kappa_{dd}$ and $\kappa_{pd}$, respectively. We also include the Fe atomic $3d$ and $2p$ spin orbit interaction (66 meV and 8.199 eV, respectively). We further include the octahedral crystal field splitting $10Dq$, as well as the strain induced, tetragonal crystal field distortion $\Delta_{e_g}\left(\equiv 2\Delta_{t_{2g}}\right)$. Also included are the charge transfer energy $\Delta$ as well as the monopole parts of the valence-valence and core-valence Coulomb interactions ($U_{dd} = 6$ eV and $U_{pd}=8$ eV, respectively). Finally, hybridization between the Fe $3d$ and ligand orbitals is included via $O_h$ symmetry hopping integrals $V_{e_g}$ and $V_{t_{2g}}\left(\equiv 0.58V_{e_g}\right)$ \cite{Robert}, and in the XAS final state the hopping integrals are rescaled by $V_f$ to account for orbital contraction due to the core hole.

The parameters used to calculate the CaFeO\textsubscript{3} x-ray absorption spectra were optimized by comparing the calculated x-ray linear dichroism to the experimental data. A single model parameter was systematically varied and all 10 independent spectra (arising from orbital and spin degeneracies of the initial state as described in the main text) were calculated for each parameter setting. The resulting x-ray linear dichroism spectra from each data set were then used to obtain the best possible fit to the experimental spectrum, illustrated here using the CaFeO\textsubscript{3}/LSAT sample. The parameter value that gave the best fit to experiment (lowest $\chi^2$ value) was taken as the optimized value. After optimizing all six model parameters in this manner, this process was repeated until the optimized parameters converged. The final parameters are shown in Table \ref{table_parameters}.

 \begin{figure}[H]
 \centering
 \includegraphics{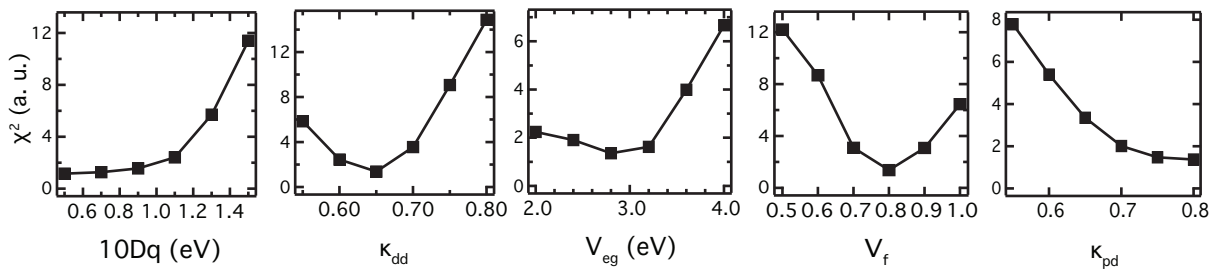}
 \caption{Goodness of fit ($\chi^2$) for the calculated x-ray linear dichroism as a function of crystal field ($10D_q$), intra-shell Coulomb interaction rescaling ($\kappa_{dd}$), hopping integrals ($V_{e_g}$, with $V_{t_{2g}}=0.58V_{e_g}$ \cite{Robert}), final state hopping rescaling ($V_f$), and inter-shell Coulomb and exchange interaction rescaling ($\kappa_{pd}$). The relevant set of 10 independent spectra was calculated for each parameter value and then the weights of each spectrum were fit to the experimental CaFeO\textsubscript{3}/LSAT x-ray linear dichroism data to obtain the $\chi^2$ value. 
 \label{SI_Jz_fitting}}
 \end{figure}

\begin{table}[H]
\squeezetable
\begin{ruledtabular}
\centering
\caption{\label{table_parameters} Optimized parameters for the ligand field simulations of the x-ray absorption and linear dichroism for CaFeO\textsubscript{3}: Crystal field ($10D_q$), hopping ($V_{eg}$), final state hopping rescaling factor ($V_f$), intra-shell Coulomb interaction rescaling factor ($\kappa_{dd}$), inter-shell Coulomb and exchange interaction rescaling factor ($\kappa_{pd}$), and charge transfer energy ($\Delta$).}
\begin{tabular}{lcccccc}
Parameter		&$10D_q$ (eV)	&$V_{eg}$ (eV)	&$V_f$	&$\kappa_{dd}$	&$\kappa_{pd}$	&$\Delta$ (eV) \\
Value		&0.5			&2.80		&0.80	&0.65			&0.80			&-2.0
\end{tabular}
\end{ruledtabular}
\end{table}

\newpage
\subsection{IV. Breakdown of the $J_z$ fits}
Tables 2 and 3 show the breakdown of the fit of the calculated x-ray linear dichroism spectrum to the experimental CaFeO\textsubscript{3}/LSAT and CaFeO\textsubscript{3}/SLAO spectra, respectively. Under tensile strain, states having preferential $x^2-y^2$ occupation are lower in energy than those having preferential $3z^2-r^2$ occupation, and their respective spin multiplicity derived spectra (labelled by $J_z$) are split in energy by a few meV due to the Fe $3d$ spin-orbit coupling. The resulting coefficient value (fit weight) and its 95\% confidence interval are shown for the optimized fit. Although the relative $J_z$ contributions do not follow a Boltzmann distribution \cite{Haverkort_LaTiO3_Jz_boltzmann}, this is not unexpected given that a Boltzmann distribution would not reproduce the inverted orbital polarization. Moreover, small distortions of the local symmetry of the crystal field are expected to change the relative energy alignment of the $J_z$ spectra, particularly given the small separation in energy of roughly 5 meV. 

To determine the $e_g$ orbital occupation, the weight of each $J_z$ spectrum is multiplied by its electron occupation in the $x^2-y^2$ and $3z^2-r^2$ orbitals. These reported occupations are the electron count in the Fe $3d$ orbitals only and do not include the ligand hole count. The resulting values are then summed for $d_{x^2-y^2}$ and $d_{3z^2-r^2}$ in order to obtain the $e_g$ electron occupation values. 

 \begingroup
\squeezetable
\begin{table}[h]
\begin{ruledtabular}
\centering
\caption{\label{}$J_z$ fit breakdown for CaFeO\textsubscript{3}/LSAT, $\Delta e_g$ = +40 meV. }
\begin{tabular}{l p{1.5cm} p{1cm} p{1.3cm} p{1cm} p{1.4cm} p{1cm} p{1cm} p{1.2cm} p{1.2cm} }
$J_z$ spectrum	&Energy above ground state (meV)	&Fit weight	&$\pm$95\% conf. int.	&Weight (\%)	&95\% conf. int. (\%)	&$x^2-y^2$ occup.	&$3z^2-r^2$ occup.	&Weight\% *occup $x^2-y^2$	&Weight\% *occup $3z^2-r^2$ \\
\hline
pref $3z^2-r^2$, $J_z = 0$	&23		&3.06	&0.57	&18.8	&3.5		&0.71	&1.14	&0.13	&0.21 \\
pref $3z^2-r^2$, $J_z = \pm 1$	&22		&5.74	&0.89	&35.3	&5.5		&0.71	&1.14	&0.25	&0.40 \\
pref $3z^2-r^2$, $J_z = \pm 2$	&18		&0.00	&0.36	&0.0		&2.2		&0.71	&1.14	&0.00	&0.00 \\
pref $x^2-y^2$, $J_z = \pm 2$	&5		&4.26	&0.46	&26.2	&2.8		&1.14	&0.71	&0.30	&0.19 \\
pref $x^2-y^2$, $J_z = \pm 1$	&2		&1.06	&0.67	&6.5		&4.1		&1.14	&0.71	&0.07	&0.05 \\
pref $x^2-y^2$, $J_z = 0$		&0		&2.15	&0.54	&13.2	&3.3		&1.14	&0.71	&0.15	&0.09 \\
\hline
Sum						&		&16.27	&		&100.0	& 		& 		&		&0.91	&0.94 \\
\end{tabular}
\end{ruledtabular}
\end{table}
\endgroup

 \begingroup
\squeezetable
\begin{table}[h]
\begin{ruledtabular}
\centering
\caption{\label{}$J_z$ fit breakdown for CaFeO\textsubscript{3}/SLAO, $\Delta e_g$ = -30 meV.}
\begin{tabular}{l p{1.5cm} p{1cm} p{1.3cm} p{1cm} p{1.4cm} p{1cm} p{1cm} p{1.2cm} p{1.2cm} }
$J_z$ spectrum	&Energy above ground state (meV)	&Fit weight	&$\pm$95\% conf. int.	&Weight (\%)	&95\% conf. int. (\%)	&$x^2-y^2$ occup.	&$3z^2-r^2$ occup.	&Weight\% *occup $x^2-y^2$	&Weight\% *occup $3z^2-r^2$ \\
\hline
pref $x^2-y^2$, $J_z = \pm$2	&20		&2.97	&0.18	&18.4	&1.1		&1.15	&0.71	&0.21	&0.13\\
pref $x^2-y^2$, $J_z = \pm$1	&17		&3.63	&0.3		&22.5	&1.9		&1.14	&0.72	&0.26	&0.16\\
pref $x^2-y^2$, $J_z = 0$		&16		&1.44	&0.16	&8.9		&1.0		&1.13	&0.72	&0.10	&0.06\\
pref $3z^2-r^2$, $J_z = 0$	&7		&1.62	&0.24	&10.1	&1.5		&0.71	&1.14	&0.07	&0.11\\
pref $3z^2-r^2$, $J_z = \pm1$	&4		&1.96	&0.26	&12.2	&1.6		&0.71	&1.14	&0.09	&0.14\\
pref $3z^2-r^2$, $J_z = \pm2$	&0		&4.49	&0.22	&27.9	&1.4		&0.71	&1.14	&0.20	&0.32\\
\hline
Sum						&		&16.11	&		&100.0	&		&		&		&0.93	&0.93\\
\end{tabular}
\end{ruledtabular}
\end{table}
\endgroup

\newpage
 \subsection{V. Anisotropic band broadening scenarios}
 
\begin{figure}[h]
 \includegraphics{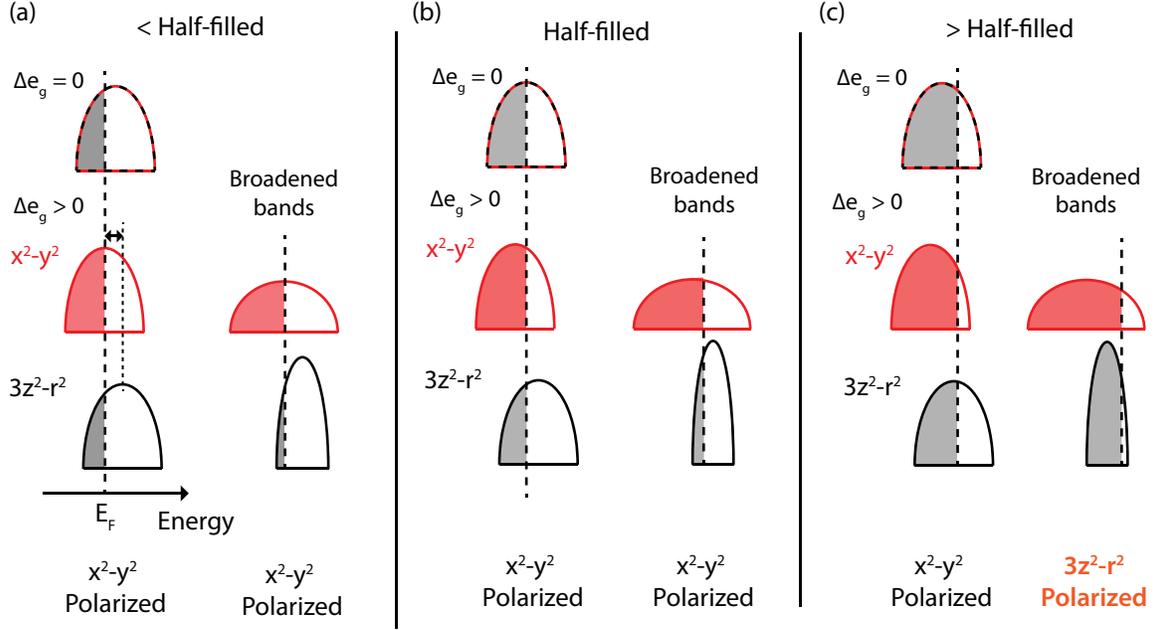}
 \caption{Simplified schematic of the effect of changes in bandwidth on the resulting orbital polarization for a system under tensile strain ($\Delta e_g>0$), which shifts the band center of masses about the unstrained center. Broadening of the in-plane band and narrowing of the out-of-plane band for the (a) less-than-half-filled band and (b) half-filled band results in the conventional preferential $x^2-y^2$ orbital polarization. (c) For a band with greater than half-filling, the same broadening can result in an inverted orbital polarization with $3z^2-r^2$ preferentially occupied.
 \label{Fig_bandwidths}}
 \end{figure}

Bandwidth effects are illustrated in Fig. \ref{Fig_bandwidths} for three scenarios: a less-than-half-filled band, a half-filled band, and a greater-than-half-filled band. In the simplest picture, the unstrained system has equal bandwidths in all three directions. Applying tensile strain ($\Delta e_g > 0$) lowers the $x^2-y^2$ band in energy relative to the $3z^2-r^2$ band and, as seen in Fig. \ref{Fig_bandwidths}(a), the $x^2-y^2$ band is more occupied. For the less-than-half-filled band (Fig. \ref{Fig_bandwidths}(a)) and the half-filled band (Fig. \ref{Fig_bandwidths}(b)), adding in the tensile strain-induced anisotropic bandwidth effects, where the in-plane $x^2-y^2$ band broadens and the out-of-plane $3z^2-r^2$ band narrows, still results in $x^2-y^2$ being more occupied and thus does not replicate our findings. 

In contrast, for a greater-than-half-filled band the band broadening can result in an inverted orbital polarization. As seen in Fig. \ref{Fig_bandwidths}(c), band broadening results in the higher energy edge of the $x^2-y^2$ band surpassing the $3z^2-r^2$ band edge, and the $3z^2-r^2$ band becomes more occupied than $x^2-y^2$. From this simple illustration, an inverted orbital polarization can be expected when the net change in bandwidth is greater than the strain-induced $\Delta e_g$ and the bands are more-than-half-filled.

\end{document}